\documentclass[%
preprint,
superscriptaddress,
amsmath, 
amssymb,
aps,
floatfix,
]{revtex4-1}
\usepackage{lipsum, babel}
\usepackage{printlen}
\usepackage{float}
\usepackage[toc,page]{appendix}
\usepackage{graphicx}% Include figure files
\usepackage{dcolumn}% Align table columns on decimal point
\usepackage{bm}% bold math
\usepackage[colorlinks=true,citecolor=blue,linkcolor=magenta]{hyperref}
\usepackage[utf8]{inputenc}
\usepackage{url}
\usepackage[normalem]{ulem}
\usepackage{upgreek}
\usepackage[mathlines]{lineno}%Enable numbering of text and display math
\usepackage{amsmath}
\usepackage{xr}
\usepackage{setspace}
\usepackage{verbatim}

\newcommand{\cmt}[1]{{}}
\newif\ifproofread

\usepackage[usenames,dvipsnames]{color}

\usepackage{xr}
\makeatletter
\newcommand*{\addFileDependency}[1]{% argument=file name and extension
  \typeout{(#1)}
  \@addtofilelist{#1}
  \IfFileExists{#1}{}{\typeout{No file #1.}}
}
\makeatother

\begin{document}
\setstretch{1.08}
\proofreadtrue

\title{Simultaneous Type-0 and Type-I Optical Parametric Oscillation in Submicron Poled Thin-Film Lithium Niobate}

\author{Fengyan Yang}
\affiliation{Department of Electrical Engineering, Yale University, New Haven, CT 06511, USA}

\author{Hong X. Tang}
\affiliation{Department of Electrical Engineering, Yale University, New Haven, CT 06511, USA}
\affiliation{Corresponding author: hong.tang@yale.edu}
\date{\today}

\begin{abstract}
\vspace{12pt}

We demonstrate dual optical parametric oscillations in a thin-film lithium niobate microring resonator enabled by polarization-insensitive submicron periodic poling. Under degenerate backward quasi-phase matching, the counter-propagating signal and idler wavevectors cancel, such that the phase-matching condition is determined solely by the pump wavevector.
As a result, a single submicron poling period simultaneously supports both Type-0 (TM $\rightarrow$ TM + TM) and Type-I (TM $\rightarrow$ TE + TE) parametric interactions in the same device. Temperature-controlled resonance alignment further enables selective activation of either polarization channel. We observe optical parametric oscillation with thresholds of 290\,{\textmu}W for the Type-0 process and 3\,mW for the Type-I process, both comparable to state-of-the-art on-chip OPO thresholds. These results establish submicron poled TFLN as a compact and reconfigurable platform for polarization-diverse parametric light generation.
\end{abstract}

%\setboolean{displaycopyright}{false} % Do not include copyright or licensing information in submission.
\maketitle
\section{Introduction}

Optical parametric oscillators (OPOs) and their below-threshold quantum counterpart, spontaneous parametric down conversion (SPDC), are versatile light sources with wide-ranging applications in both the classical and quantum regimes, such as generating laser light with frequencies beyond the reach of conventional optical gain media, spectroscopy \cite{CW_pulse_OPO,spec_5_12um,spec_infrared}, quantum computing \cite{Ising_DOPO,Ising_network}, and advanced metrology \cite{review_squeezed,amir_squeezed,QKD,squeeze_NC,2013_ligo_squeeze}. In integrated photonics, thin-film lithium niobate (TFLN) has emerged as a leading platform for realizing compact and efficient $\chi^{(2)}$ devices, thanks to its large nonlinear coefficient, low propagation loss, and ferroelectric property for domain engineering \cite{cheng_2018_efficient,29mQ_LN,Mfejer1995,Zhu2021}. TFLN microring resonators in particular have demonstrated several $\chi^{(2)}$-OPOs with low-threshold \cite{Juanjuan_OPO,OPO_amir,Bruch2019,Kellner2025}, and broadband tunability \cite{Ledezma2023}.

However, most OPOs to date are designed to support a single nonlinear process, which is typically co-propagating Type-0 or Type-I phase matching, limiting the flexibility and reconfigurability of these parametric sources. In general, achieving different types of OPOs (Type-0, I, or II) requires corresponding phase-matching engineering for interactions based on different nonlinear tensor elements. In bulk crystals, this is typically accomplished via birefringent phase matching through crystal angle tuning (e.g., in BBO or LBO) \cite{typeII_BBO1995,OPO_BBO_LBO}, while in ferroelectric thin-film materials like KTP and lithium niobate, periodic poling allows quasi-phase matching of specific polarization and frequency combinations \cite{Arbore1997}. Simultaneous realization of multiple nonlinear processes in a single structure has been explored in both bulk and chip-scale systems, including approaches that use spatial-mode engineering, composite periodic poling gratings, or higher-order QPM \cite{Linran_typeIandII,Gao_MBPM_spdc,Luo2019}, though these approaches often face trade-offs in efficiency, complexity and controllability.

%SPDC, as the below-threshold counterpart of OPO, further motivates the pursuit of simultaneous backward Type-0 and Type-I interactions for quantum light generation. Such a scheme would provide counter-propagating photon pairs with spatial separation and polarization diversity, enabling heralded single-photon sources and hybrid entangled states%This capability is particularly attractive for integrated quantum photonic circuits requiring efficient photon routing and filtering. \cite{MYG_polarization_entanglement,lu2025counterSPDC_GaSe, Qsplitter}. Recent demonstrations of counter-propagating SPDC underscore the potential of submicron ferroelectric patterning for integrated quantum light sources based on backward $\chi^{(2)}$ processes \cite{Sabatti2025, Kellner2026CP}.

In this work, we realize both Type-0 and Type-I optical parametric oscillation in a single submicron-poled thin-film lithium niobate microring. Owing to degenerate backward phase matching \cite{fengyan_SSHG, DLOPO}, the signal and idler wavevectors cancel, so a single poling period can phase match both TM $\rightarrow$ TM + TM and TM $\rightarrow$ TE + TE interactions. We experimentally observe low-threshold oscillation in both channels and thermally switch between them through resonance alignment. This work introduces a compact, reconfigurable nonlinear platform that supports polarization-diverse parametric light generation, paving the way for integrated sources with both classical and quantum functionality.

\section{Design and Fabrication of Microring}
\begin{figure*}[h]
\centering
\includegraphics[width=0.75\linewidth]{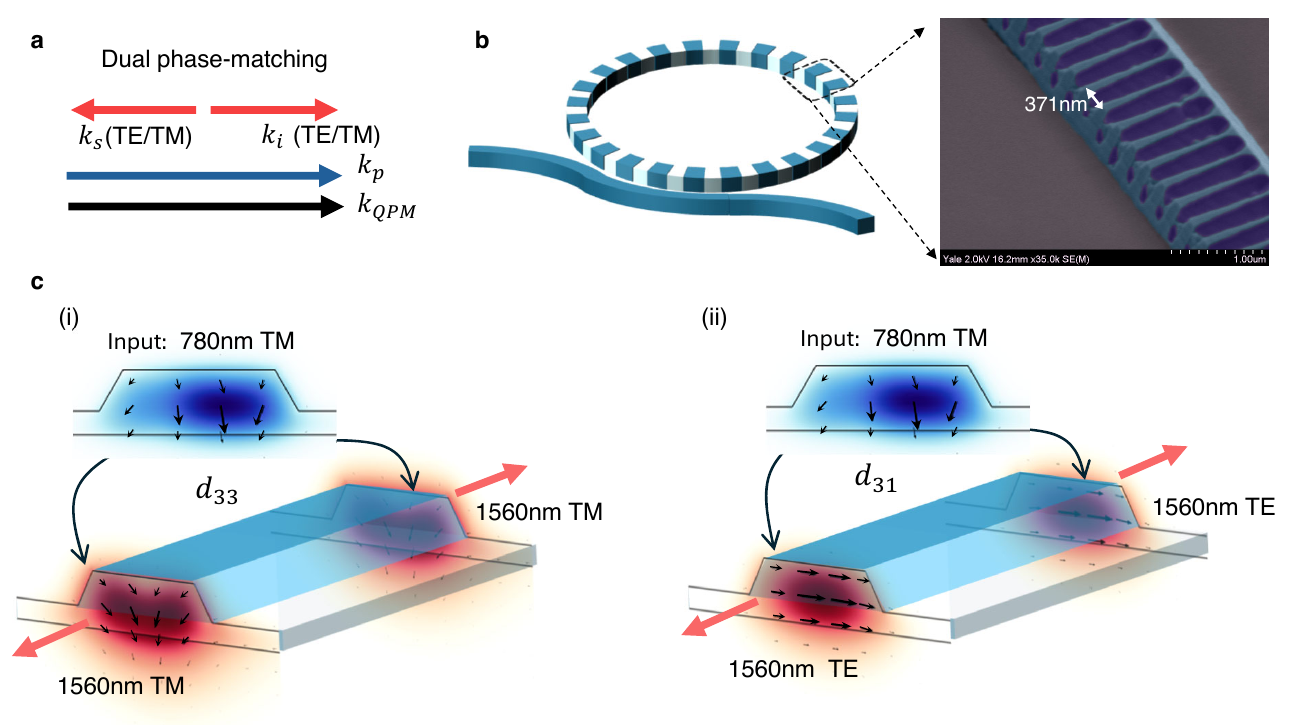}
\caption{(a) Lithium niobate ring resonator with 371\,nm poling period. (b) Dual backward phase-matching condition. (c) Type-0 (i) and Type-I (ii) optical parametric oscillation in the backward phase-matched scheme.}
\label{fig1_principle}
\end{figure*}

The operating principle of our device is illustrated in Fig.~\ref{fig1_principle}(a). Backward phase matching at degeneracy enforces a counter-propagating configuration in which the signal and idler wavevectors ($k_s$ and $k_i$) are equal in magnitude and opposite in direction. Under this condition, their contributions cancel in the momentum conservation relation, and the quasi-phase-matching requirement reduces to $k_p - k_{\mathrm{QPM}} = 0$, meaning that the poling wavevector $k_{\mathrm{QPM}}$ compensates only the pump momentum. This key simplification enables a single poling period to satisfy multiple polarization configurations. Figure~\ref{fig1_principle}(c) also shows the corresponding interaction schemes: a 780\,nm TM$_{00}$ mode interacts with either the 1560\,nm TM$_{00}$ mode through the $d_{33}$ tensor element (Type-0 process) or the 1560\,nm TE$_{00}$ mode through the $d_{31}$ tensor element (Type-I process). Because both processes share the same 780\,nm wavevector at degeneracy, they require the same poling period for quasi-phase-matching, which is calculated to be $\Lambda=\lambda_{vis}/n_{\mathrm{eff,vis}}=$371\,nm. The false-color scanning electron microscope (SEM) image in Fig.~\ref{fig1_principle}(b) confirms the realization of submicron periodic poling in the microring, where the inverted domains are selectively etched by hydrofluoric acid and appear as trenches marked by dark blue in the SEM image. Together, these features establish the physical basis for simultaneous Type-0 and Type-I parametric oscillation within a single backward-phase-matched TFLN resonator.

\begin{figure*}[h]
\centering
\includegraphics[width=1\linewidth]{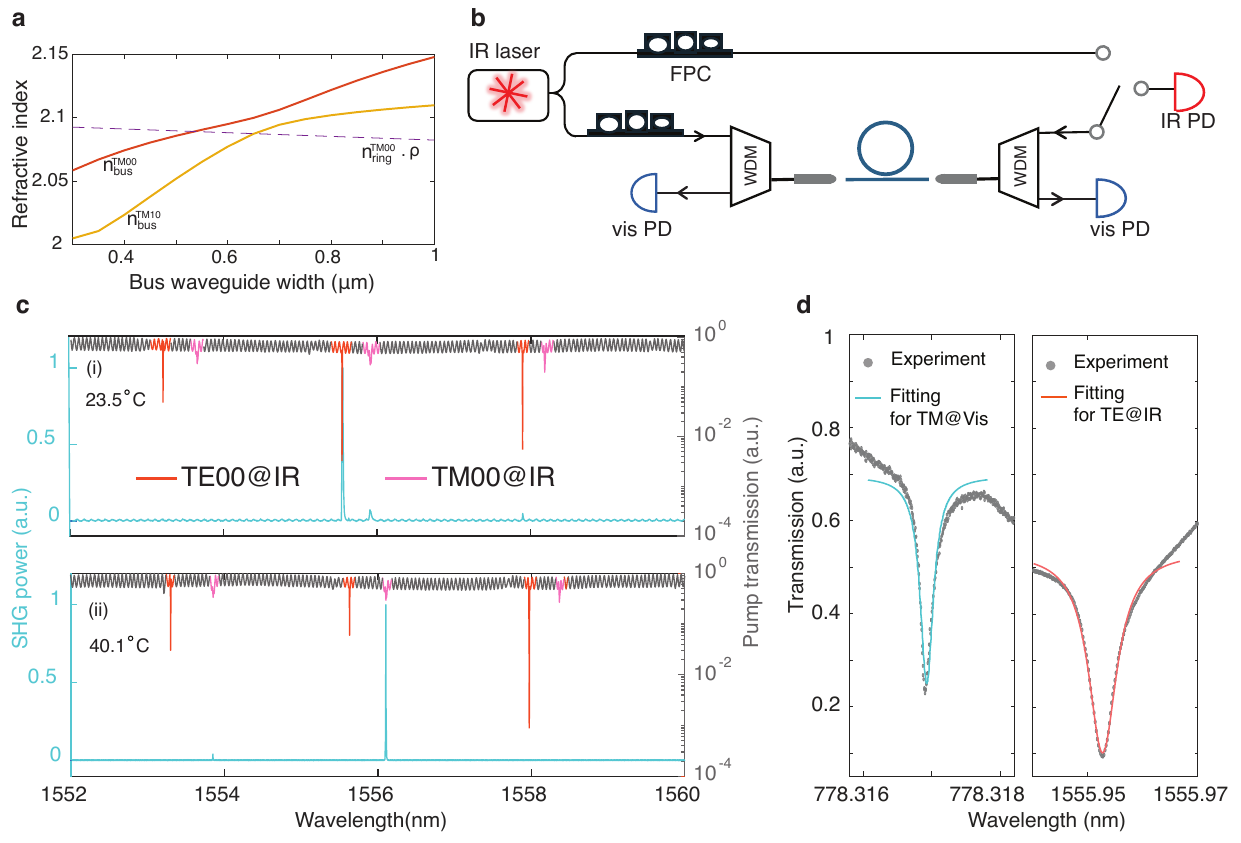}
\caption{
(a) Index matching sweep as bus width changes. $\rho=\frac{R+W_{\mathrm{ring}}/4}{R+W_{\mathrm{gap}}+W_{\mathrm{ring}}/2+W_{\mathrm{bus}}/2}$ is the effective radius ratio between the ring mode and bus mode. 
(b) Experimental setup for simultaneous measurement of the 1560 nm transmission and the symmetric second-harmonic generation at 780 nm. IR PD: infrared photodetector; VIS PD: near-visible photodetector; WDM: 780 nm/1560 nm wavelength demultiplexer; FPC: fiber polarization controller.
(c) Temperature-dependent SHG spectra overlaid with the corresponding infrared transmission.
(i) At 23.5$^\circ$C, the SHG peak aligns with the 1560\,nm TE$_{00}$ resonance, corresponding to the Type-I SHG process.
(ii) At 40.1$^\circ$C, the SHG peak aligns with the 1560\,nm TM$_{00}$ resonance, corresponding to the Type-0 SHG process.
(d) Measured cavity resonance lineshapes and Lorentzian fittings for the 780\,nm TM$_{00}$ mode (left) and the 1560\,nm TE$_{00}$ mode (right), from which the loaded quality factors are extracted.
}
\label{fig2_spectrum_Qfitting}
\end{figure*}

We designed the microring to support simultaneous coupling of the 780\,nm pump in TM polarization and the 1560\,nm signal/idler bands in both TE and TM polarizations. In cavity-based OPOs, there is an inherent trade-off between oscillation threshold and conversion efficiency: reducing the threshold generally favors a higher loaded $Q$ and therefore under coupling, whereas efficient extraction of the generated fields requires sufficiently over-coupling \cite{Juanjuan_OPO}. In this work, we therefore target a compromise between maintaining a high loaded $Q$ for the pump and enabling efficient out-coupling of the signal and idler fields.

To realize this across widely separated wavelengths, we adopt a wraparound (pulley) coupler geometry, which provides additional freedom for engineering the coupling strength simultaneously at 780\,nm and 1560\,nm \cite{wraparound,pulley}. Coupling between the bus waveguide and the ring is governed by both modal overlap and phase matching. Because the 780\,nm mode is more tightly confined and exhibits weaker evanescent overlap, its phase-matching condition is more restrictive. We therefore prioritize phase matching at the pump wavelength when determining the bus geometry. Specifically, we fix the ring radius at 70\,{\textmu}m, the ring width at 1.4\,{\textmu}m, the coupling gap at 0.4\,{\textmu}m, and an etch depth of 0.4\,{\textmu}m in a 600\,nm z-cut lithium niobate thin film. We then sweep the bus-waveguide width to identify the phase-matched coupling region, as shown in Fig.~\ref{fig2_spectrum_Qfitting}(a). From this sweep, a width of 0.53\,{\textmu}m is selected, which provides best phase matching for the pump. The wraparound angle is determined to be 40 degrees for stronger coupling at 1560\,nm.

For nonlinear frequency conversion, periodic poling with a 371\,nm period was implemented by applying voltage to nickel electrodes patterned on top of the TFLN film, followed by patterning of the microring resonators within the poled area. For additional fabrication details, please see \cite{fengyan_SSHG}. We also applied high-$Q$ fabrication strategies, including shallow etching to minimize mode overlap with the sidewall and optimized wet etching conditions (RCA solution at 50\,$^\circ$C for 10\,min) to reduce defects in the inverted domains.

To characterize the fabricated devices, we measured the loaded quality factors ($Q$) for both the 780\,nm TM mode and the 1560\,nm TE mode. The 780\,nm TM pump mode exhibits an intrinsic quality factor of $Q_{0}=4.2\times10^{6}$ and a loaded quality factor of $Q_{\mathrm{L}}=3.3\times10^{6}$, while the 1560\,nm TE mode exhibits $Q_{0}=1.4\times10^{6}$ and $Q_{\mathrm{L}}=5\times10^{5}$, as presented in Fig.\,\ref{fig2_spectrum_Qfitting}(d). The 1560\,nm TM mode 
did not show a well-defined Lorentzian lineshape suitable for reliable fitting, likely due to stronger overcoupling. Nevertheless, by estimating the linewidth of the split resonance, we obtain $Q_L\simeq1.2\times10^{5}$ and $Q_{\text{in}}\simeq6.5\times10^{5}$ for the 1560\,nm TM mode.

%The experimental loaded $Q$ for the 780\,nm TM mode is $3.4 \times 10^6$ and , while the 1560\,nm TE mode exhibits a loaded $Q$ of about $0.5 \times 10^6$, as presented in Fig.\,\ref{fig2_spectrum_Qfitting}(d). The 1560\,nm TM mode did not show a well-defined lineshape suitable for fitting, which we attribute to stronger overcoupling.

%To confirm the coupling regime more rigorously, we performed single-sideband modulation (SSBM) assisted measurements on the 1560\,nm TE mode. This technique down-converts the optical resonance response to the microwave domain, enabling direct measurement of the phase shift across the resonance. By analyzing this phase behavior, we confirmed that the 1560\,nm TE mode is indeed over-coupled. From this analysis, we extracted an intrinsic $Q$ of approximately $1.8 \times 10^6$ \added{and} a coupling $Q$ of $0.7 \times 10^6$ for the TE mode. 

\section{Measurement results }
Before measuring OPO directly, we first use the corresponding symmetric second harmonic generation (SSHG) processes as a convenient probe to identify the two nonlinear conversion channels and calibrate their temperature-dependent switching behavior. Both the Type-0 (1560\,nm\,TM\,$\rightarrow$\,780\,nm\,TM) and Type-I (1560\,nm\,TE\,$\rightarrow$\,780\,nm\,TM) SSHG processes in our device are designed to be phase matched through the same periodic poling structure. Which process dominates experimentally depends on the resonance alignment condition, $\delta_\omega(T) = 2\,\omega_{\mathrm{IR}}(T) - \omega_{\mathrm{VIS}}(T),$ which can be tuned via temperature $T$. At a given temperature, if $\delta_\omega^{\mathrm{TM}} \ll \delta_\omega^{\mathrm{TE}}$, then the Type-0 process will dominate. In other words, each process has its own optimal operating temperature and we can selectively activate one over the other through temperature tuning.

To experimentally characterize this behavior, we performed temperature-tuning measurements. Figure~\ref{fig2_spectrum_Qfitting}(b) illustrates the experimental setup. A continuous-wave 1560\,nm laser is injected into the device from one end of the coupling waveguide. Due to reflection at the opposite facet, a weak counter-propagating 1560\,nm field is also present inside the ring, effectively fulfilling the requirement for symmetric SHG. In this configuration, when the optical switch is set to the lower arm connected to the infrared (IR) photodetector, we can simultaneously monitor the 1560\,nm transmission and the corresponding 780\,nm SHG signal.

Fig.\,\ref{fig2_spectrum_Qfitting}(c) demonstrates how temperature tunes the resonance matching between the fundamental and the second-harmonic modes. At 23.5\,$^\circ$C, the SHG peak [Fig.\,\ref{fig2_spectrum_Qfitting}(c-i)] aligns well with the 1560\,nm TE resonance, indicating optimal conditions for the Type-I process. When the temperature is increased to 40.1\,$^\circ$C, the SHG peak shifts to align with the TM mode, favoring the Type-0 process.

\begin{figure*}[t]
\centering
\includegraphics[width=1\linewidth]{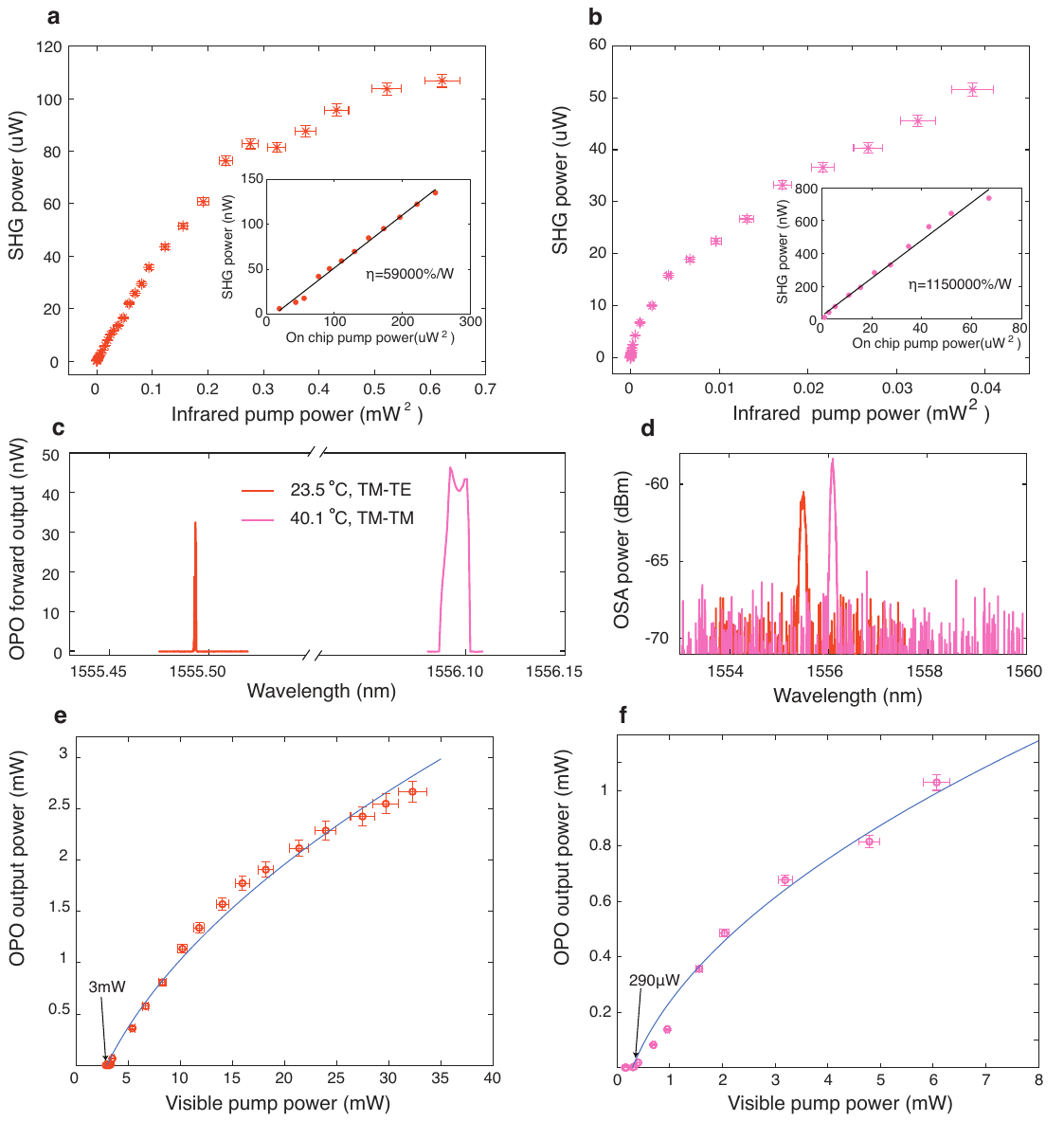}
\caption{
(a-b) Measured Type-I (Type-0) symmetric SHG efficiency.
(c) OPO output power from one 1560\,nm port as the visible pump wavelength is tuned across the resonance, measured at the optimized temperatures for the Type-I and Type-0 processes.
(d) OSA spectra of the single-side OPO output for the Type-I and Type-0 processes, with the same color coding as in (c).
(e) OPO output power versus on-chip pump power for the Type-I process, showing a threshold of approximately 3\,mW. 
(f) OPO output power versus on-chip pump power for the Type-0 process, showing a threshold of 290\,{\textmu}W.
}
\label{fig3SHGOPO}
\end{figure*}

To quantify the SHG efficiency, we further tuned to the optimal temperature for each process and activated the symmetric pump configuration (switch in [Fig.\,\ref{fig2_spectrum_Qfitting}(b)] connected to the upper port) to measure the generated SHG power. The results are shown in Fig.\,\ref{fig3SHGOPO}(a,b). In the non-depleted regime, the extracted normalized conversion efficiency for the Type-I SSHG is $5.89 \times 10^4$\%/W, while for the Type-0 SSHG, it reaches $1.15 \times 10^6$\%/W. The observed efficiency ratio is consistent with the relative magnitudes of the nonlinear tensor components involved. With $d_{33} \approx 19.6\,\mathrm{pm/V}$ and $d_{31} \approx 3.2\,\mathrm{pm/V}$, the expected scaling $\eta_{\mathrm{SHG}} \propto d_{\mathrm{eff}}^2$ gives $(d_{33}/d_{31})^2 \sim 37$, reasonably consistent with experiment ratio of $\sim$ 20, with the remaining discrepancy attributable to differences in coupling conditions and cavity loaded $Q$ factors.

To measure the OPO response, a visible pump laser is injected into the 780\,nm port of the left WDM, and the counter-propagating signal and idler fields generated through parametric oscillation emerge from both 1560\,nm output ports of the WDM and are collected using two infrared photon detectors. This configuration naturally achieves degenerate operation, in which the signal and idler frequencies are automatically locked to the degenerate point through the counter-propagating phase-matching condition \cite{DLOPO}. The same device supports both the Type-0 (780\,nm\,TM\,$\rightarrow$\,1560\,nm\,TM + TM) and Type-I (780\,nm\,TM\,$\rightarrow$\,1560\,nm\,TE + TE) parametric oscillation processes. Each process can be selectively activated by adjusting the device temperature to align the infrared and visible resonances.

We further characterize the OPO emission by tuning the visible pump wavelength across the resonance and recording the generated infrared output from one WDM output port, as shown in Fig\,\ref{fig3SHGOPO}(c). The OPO peaks appear at the corresponding degenerate wavelengths for the two operating temperatures, confirming selective activation of the Type-I and Type-0 processes. The broader feature of the Type-0 process in Fig\,\ref{fig3SHGOPO}(c) reflects a larger pump-detuning range for oscillation. This is expected because the 1560 nm TM mode is more strongly overcoupled and has a lower loaded Q, giving a broader resonance linewidth. Together with the stronger $d_{33}$-mediated nonlinear gain, the Type-0 OPO can remain above threshold over a wider pump-wavelength range. The optical spectra measured by the OSA [Fig\,\ref{fig3SHGOPO}(d) ] show OPO emission for both processes. For conciseness, Fig\,\ref{fig3SHGOPO}(c,d) show the single-side OPO output.

Fig\,\ref{fig3SHGOPO}(e-f) show the measured total degenerate OPO output power collected from both
1560 nm ports as a function of on-chip pump power for the two different processes, while the
solid curves represent analytical fittings. For the TM-TE (Type-I) process [Fig\,\ref{fig3SHGOPO}(e)], we
observe clear threshold behavior with a measured OPO threshold of approximately 3 mW, with a maximum conversion efficiency of 11\%. For the TM-TM (Type-0) process [Fig\,\ref{fig3SHGOPO}(f)],
the threshold is significantly lower at around 290~$\mu$W while the maximum conversion efficiency reaches 23\% thanks to the stronger over-coupling of 1560 nm TM mode. Using the measured SHG efficiencies together with the extracted cavity quality factors, we deduce effective coupling strengths of $g_{31}/2\pi=0.026$\,MHz and $g_{33}/2\pi=0.37$\,MHz, corresponding to predicted OPO thresholds of 2.6\,mW for the Type-I process and 226\,$\mu$W for the Type-0 process, respectively, in good agreement with the measured values.

These results confirm the device’s ability to achieve low-threshold OPO operation with decent conversion efficiency in both Type-0 and Type-I configurations using the same submicron-poled ring resonator. The combination of backward phase matching, submicron periodic poling, and high-quality factor resonator design enables flexible, efficient parametric oscillation across both polarization channels.

\section{Discussion and Conclusion}

In summary, we demonstrate a submicron-periodically poled thin-film lithium niobate microring resonator that supports efficient, low-threshold optical parametric oscillation in both Type-0 and Type-I configurations. By leveraging submicron-scale periodic poling for backward phase matching, the device enables degenerate OPO with distinct polarization combinations, which can be selectively accessed via temperature tuning. These results establish submicron-poled thin-film lithium niobate microrings as a versatile platform for integrated, tunable, and polarization-diverse parametric sources.

Beyond classical operation, the simultaneous availability of backward Type-0 and Type-I interactions is also relevant for quantum photonics, as their below-threshold SPDC counterparts could generate counter-propagating photon pairs with both polarizations and spatial separation, facilitating routing and filtering in integrated circuits \cite{MYG_polarization_entanglement,lu2025counterSPDC_GaSe}. Recent demonstrations of counter-propagating SPDC further highlight the potential of submicron ferroelectric patterning for integrated backward $\chi^{(2)}$ light sources \cite{Sabatti2025,Kellner2026CP}.

More broadly, the ability to access multiple parametric processes within a single device provides a pathway toward compact and reconfigurable nonlinear photonic systems. This capability may enable multifunctional frequency conversion, polarization-engineered parametric sources, and scalable architectures for integrated nonlinear and quantum photonics. Future work will focus on extending these results below threshold to realize chip-scale sources of entangled photon pairs or squeezed states.

\smallskip{}

%\begin{backmatter}
\section*{Disclosures}
The authors declare no conflicts of interest.
\section*{Acknowledgments}
This work is supported by DARPA through its INSPIRED program under cooperative agreement D24AC00180. The part of the research that involves lithium niobate thin film preparation is supported by the US Department of Energy Co-design Center for Quantum Advantage (C2QA) under Contract No. DE-SC0012704. HXT acknowledges support from the National Science Foundation through Award No 2410725. 

The authors would like to thank Yong Sun, Lauren Mccabe, Kelly Woods, and Michael Rooks for their assistance provided in the device fabrication. The fabrication of the devices was done at the Yale School of Engineering \& Applied Science (SEAS) Cleanroom and the Yale Institute for Nanoscience and Quantum Engineering (YINQE).

%\bmsection{Supplemental document}
%See Supplement 1 for supporting content.

%\end{backmatter}

\bibliography{main}

\end{document}